\documentstyle[aps,prb,epsf]{revtex}

\newcommand{\ba}{\begin{eqnarray}}
\newcommand{\be}{\begin{equation}}
\newcommand{\ea}{\end{eqnarray}}
\newcommand{\ee}{\end{equation}}
\newcommand{\up}{\uparrow}
\newcommand{\down}{\downarrow}

\begin{document}

\draft

%%% REMOVE FOR SUBMISSION
\twocolumn[\hsize\textwidth\columnwidth\hsize\csname@twocolumnfalse%
\endcsname

\title{Schwinger boson theory of anisotropic ferromagnetic
ultrathin films}
\author{Carsten Timm$^{1}$\cite{email} and
P.J. Jensen$^{2,1}$}
\address{$^1$Institut f\"ur Theoretische Physik, Freie Universit\"at
Berlin, Arnimallee 14, D-14195 Berlin, Germany\\
$^2$Hahn-Meitner-Institut Berlin, Glienicker Str.\ 100,
D-14109 Berlin, Germany}
\date{December 21, 1999}

\maketitle

\begin{abstract}
Ferromagnetic thin films with magnetic single-ion anisotropies are studied
within the framework of Schwinger bosonization of a quantum Heisenberg
model. Two alternative bosonizations are discussed. We show that
qualitatively correct results are obtained even at the
mean-field level of the theory, similar to Schwinger boson results for
other magnetic systems. In particular,
the Mermin-Wagner theorem is satisfied: a spontaneous
magnetization at finite temperatures
is not found if the ground state of the anisotropic system
exhibits a continuous degeneracy. We calculate
the magnetization and effective anisotropies as functions of
exchange interaction, magnetic anisotropies,
external magnetic field, and temperature for arbitrary
values of the spin quantum number. Magnetic reorientation
transitions and effective anisotropies
are discussed. The results obtained by Schwinger boson
mean-field theory are compared with the many-body
Green's-function technique.
\end{abstract}

\pacs{PACS numbers: 75.10.Jm, 75.30.Gw, 75.70.Ak}

]

\narrowtext

\section{Introduction}
\label{sec:intro}

Schwinger boson theories are very successful in describing
magnetism in various quantum
systems.\cite{Schwinger,RN,AA,Auerbach,Starykh,RS,TGHS}
Qualitatively correct results are mostly obtained even in the mean-field
approximation, {\it e.g.}, quantum Heisenberg ferromagnets and
antiferromagnets are well described for arbitrary spin
quantum number for two and more spatial dimensions. In particular
in two-dimensional (2D) isotropic Heisenberg magnets no spontaneous
magnetization is found for non-zero temperatures,\cite{AA,Auerbach,RS}
in accordance with the Mermin-Wagner theorem.\cite{MW}
There are basically two reasons why Schwinger boson mean-field
(SBMF) theory works well
even in low dimensions:
(a) Since the bosonized spin degrees of freedom are integrated
over in the functional integral,
spin fluctuations are taken into account and (b) the approach
does not constitute an expansion around an ordered state, unlike
Holstein-Primakov bosonization,\cite{HP} and thus works for
both ordered and disordered ground states.

The {\it first aim\/} of the present paper is to test the applicability of
SBMF theory to another class of quantum-magnetic models, namely
the Heisenberg ferromagnet with single-ion (on-site) anisotropies.
We will restrict ourselves to the two-dimensional case, which is
the most interesting one, but the same methods can be used in any
number of dimensions.
The anisotropic Heisenberg model in two dimensions is of particular
interest, since it describes ultra-thin ferromagnetic films, including the
important cases of the $3d$ transition metals iron, cobalt, and nickel. Thin
ferromagnetic films are of great technological relevance, {\it e.g.}, for
magnetic disk drives. They also show many interesting physical effects, of
which we only mention the observed reorientation transitions as a function
of both film thickness and
temperature.\cite{ex2a,ex2b,ex2c,ex2d,ex2e,FMR,Bab97,ex2f,ex2g,ex2h,ex2i}
Much theoretical work has already been done on anisotropic films, which we
briefly review below. However, it would be desirable to have an alternative
approach for comparison. The {\it second aim\/} of this paper is to provide
such an approach using Schwinger bosonization. We apply this method to the
calculation of the magnetization with and without an external magnetic
field. In addition we determine effective anisotropies, which are crucial
for comparisons of measurements performed at finite temperatures $T$ to {\it
ab-initio\/} calculations of anisotropies at $T=0$. The effective
anisotropies describe the dependence of the free energy
on the polar angles $(\theta,\phi)$ of the magnetization.

From the theoretical point of view, two-di\-men\-sio\-nal anisotropic
ferromagnets are of particular interest, since the anisotropies reduce the
symmetry of the system. For example, an easy axis leads to a discrete,
Ising-type symmetry, while a easy plane leads to an $XY$-type symmetry.
According to the Mermin-Wagner theorem,\cite{MW} a spontaneous magnetization
at finite temperatures is possible in the first case but not in the
second.\cite{Kit51} We will show that SBMF theory yields qualitatively
correct results for both of these cases.

We now summarize previous work on thin magnetic films.
So far, most approaches rely on simple mean-field approximations,
which are of limited validity for
low-dimensional systems. In particular, they do not satisfy the
Mermin-Wagner theorem. A finite magnetization
can be induced by magnetic anisotropies,\cite{Kit51} which, however, cannot
be described by a perturbative expansion around the symmetric state.
Thus, improved approaches are needed, which should also allow to
determine the magnetic properties in the whole range
of temperatures and for systems with nonequivalent lattice sites.
For example, magnetic reorientation transitions have been studied within
spin-wave theories. In particular, Holstein-Primakov bosonization\cite{HP}
has been applied,\cite{Mil91} which is, however, limited to low
temperatures.

The magnetic properties of 2D anisotropic ferromagnets have also been
studied using many-body Green's functions, which allow
calculations in the whole temperature range of
interest.\cite{Tya59,Tha62,Cal63,Tya67,FJK99}
This method has also been applied to thin
films of several layers.\cite{Doe64,Hau72,Die79,Chi90,Bru91,Mor94}
Within this theory the higher-order Green's functions are approximated
by the so-called Tyablicov (or random-phase approximation)
decoupling,\cite{Tya59}
{\it i.e.}, one of the spin operator components is replaced
by its expectation value, whereas the expectation values of the other
two components vanishes. Note that within this approach the spin commutation
relations are preserved. By comparison with a recent Quantum-Monte-Carlo
calculation for an {\it isotropic\/} 2D
magnet with $S=1/2$ on a square lattice in an external
magnetic field,\cite{HSTG,TGHS} it has been shown that this approach
yields a reasonable description for the magnetization.\cite{EFJK} In
addition, the long-range magnetic dipole interaction can be treated as well
within the same decoupling scheme. On the other hand, {\it single-ion\/}
anisotropies have to be treated differently, since the inclusion of such
terms leads to an ambigous description.\cite{Erd75} Anderson and
Callen,\cite{And64} as well as Lines\cite{Lin67} have proposed interpolation
schemes, which are valid only for small single-ion interactions as compared
to the exchange coupling. For most cases of interest this condition is
fulfilled. In order to calculate the effective anisotropies ${\cal K}_i(T)$
and the temperature-dependent magnetic reorientation, non-vanishing
expectation values of at least two spin operator components have to be
determined simultaneously.\cite{FJK99} Whereas the inclusion of the magnetic
dipole coupling is straightforward, the inclusion of single-ion anisotropies
turns out to be tedious, in particular for spin quantum numbers $S>1$, and
for higher-order anisotropies. We emphasize that within these Green's-function
approaches the magnetic anisotropies cannot be handled as small
perturbations. Also, this approach suffers from the ambigous decoupling
schemes for these single-ion terms. Thus, a theory not suffering from these
problems would be desirable.

Finally, we summarize other approaches. First, mean-field
theories have been used,\cite{Jen96,Mil95,Jen98,JeB98,Mos94,Huc97}
which are particularly simple to apply for complicated
systems, {\it e.g.}, with nonequivalent atomic layers
in a thin film or several non-equivalent lattice sites. Classical
as well as quantum-mechanical Heisenberg spins have been studied
allowing for an arbitrary direction of the magnetization. The inclusion
of single-ion anisotropies is quite simple. A perturbative treatment of
the anisotropies yields analytical
expressions for the effective anisotropies.\cite{Jen93,JeB98,Mil95}
However, these approaches completely neglect both thermal and
quantum fluctuations, which are crucial for low-dimensional systems.
Kawazoe {\it et al.}\cite{cont} have treated the magnetic reorientation of
thin ferromagnetic films within a continuum approach for the system and
mean-field-like expressions for the magnetization density.
A continuum approach to the magnetic reorientation has also been used within
a renormalization-group treatment.\cite{renorm}

On the other hand, following earlier work\cite{Cas59,Bab97}
Millev {\it et al.}\cite{MOK} treat the magnetization of a thin film as
a macroscopic, classical magnetic moment and discuss its
ground state as a function of anisotropy and external magnetic field.
The anisotropy constants appearing in their work are to be understood
as effective anisotropies obtained from experiment or from some
more advanced calculation. This approach has to assume that the effects
of thermal and quantum fluctuations, the lattice structure, and
the spin quantum number $S$ can be described by a suitable renormalization
of the anisotropy constants. From a theoretical point of view, the
problem has just been transferred to the calculation of these effective
anisotropies. This has to be done using more advanced methods such as
Schwinger bosonization (discussed below) or
many-body Green's functions.

Finally, Monte-Carlo simulations have been performed for the
effective anisotropies and magnetic reorientation of ferromagnetic
monolayers. So far, only classical spins have been studied for
the anisotropic case,\cite{MC} while Quantum-Monte-Carlo results
exist for the isotropic case only.\cite{HSTG,TGHS} More
complicated systems such as thin films with several layers 
or including the dipole interaction
have not been studied yet.

Following the goals set above, we derive in Sec.~\ref{sec:deriv} SBMF
theories for films with second- and fourth-order single-ion anisotropy. We
employ two different bosonization schemes, known as SU($N$) and O($N$)
bosonization. In Sec.~\ref{sec:res} we show representative results and
compare them with results from simple mean-field theories and from many-body
Green's-function methods. We focus on two areas of current interest: In
Sec.~\ref{sus:1} we discuss the change of magnetization as a function of
external field and of temperature, in particular the reorientation
transitions, and present a number of phase diagrams. In Sec.~\ref{sus:2} we
calculate temperature-dependent effective anisotropies. A summary and
conclusions are given in Sec.~\ref{sec:fin}.

\section{Schwinger boson theory}
\label{sec:deriv}

In this section we summarize the SBMF theory for an anisotropic 2D
ferromagnet, using two different, but related, bosonizations. The first
one employs the SU(2) symmetry of the spins\cite{Schwinger} in the absence
of anisotropies and leads to an SU($N$)-symmetric mean-field theory, where
$N\to\infty$. The second makes use of the mapping between the Lie groups
SU(2) and O(3) and results in an O($N$) mean-field theory.\cite{RS} We first
sketch the SU($N$) theory, which is described in greater detail in
App.~\ref{app:a}. The derivation for the O($N$) theory is similar and we
only present the results.

The main idea of SU(2) Schwinger boson theory is to map the spin
operators onto boson operators $b_\up$ and $b_\down$ according
to\cite{Schwinger,AA}
\ba
S^+ & = & b_\up^\dagger b_\down ,
\label{1.S1}\\
S^- & = & b_\down^\dagger b_\up ,
\label{1.S2}\\
S^z & = & (b_\up^\dagger b_\up-b_\down^\dagger b_\down)/2 .
\label{1.S3}
\ea
This mapping is {\it exact\/}; in particular the spin commutation
relations are correctly reproduced. However, the bosonization introduces
unphysical degrees of freedom, which have to be removed by the constraint
\be
b_\up^\dagger b_\up+b_\down^\dagger b_\down= 2S ,
\label{1.Sc1}
\ee
where $S$ is the spin quantum number.
Whereas the above scheme employs the SU(2) spin symmetry group, one
can also take advantage of the mapping between SU(2) and the group
O(3)---both have the same algebra up to the choice of
representation---to formulate an alternative O(3) Schwinger boson
theory.\cite{RS}
%%INCORRECT:
%%We stress that O($3$) bosonization is in fact {\it not\/} exact.
%%To our knowledge this point has not been discussed before.
%%The spin commutators are reproduced correctly but the total spin is not:
%%${\bf S}\cdot{\bf S}$ is mapped under O(3) bosonization to $S(S+2)$
%%instead of the exact value $S(S+1)$. This result suggests that quantum
%%fluctuations are overestimated in O($N$) theory, leading to
%%unphysical behavior at low temperatures, which is indeed seen.\cite{TGHS}
%%SU($N$) theory does not suffer from this problem.
For both groups well-defined expansions around a
mean-field theory are obtained by generalization to $N$ boson fields
and expanding in small $1/N$.\cite{Auerbach} In particular, in the limit
$N\to\infty$ mean-field theory becomes exact.

Read and Sachdev\cite{RS} have derived SU($N$) and O($N$) SBMF expressions
for the magnetization of isotropic 2D ferromagnets, where their main
interest was in the quantum Hall ferromagnet at filling factor
$\nu=1$.\cite{QHferro1,QHferro2} Recently, the leading-order fluctuation
corrections, {\it i.e.}, the first order of the $1/N$ expansion, have been
calculated\cite{TGHS} and compared to Quantum-Monte-Carlo
simulations\cite{HSTG,TGHS} and experimental results.\cite{Manfra} The
SU($N$) and O($N$) theories are found to be qualitatively correct. In the
quantum Hall system the magnetic degrees of freedom are carried by
conduction electrons with $S=1/2$. Thus single-ion anisotropies do not play
any role.\cite{rem:QHdip}

We describe the anisotropic 2D ferromagnet by a Heisenberg model on a 2D
square lattice, augmented by second- and fourth-order
single-ion anisotropies,
\ba
H & = & -J \sum_{\langle ij\rangle} {\bf S}_i\cdot{\bf S}_j
  - {\bf B}\cdot \sum_i {\bf S}_i
  - K_2 \sum_i (S_i^z)^2
  \nonumber \\
& & {}- K_4^\perp \sum_i (S_i^z)^4
  - K_4^\| \sum_i \Big[(S_i^x)^4+(S_i^y)^4\Big] ,
\label{2.H1}
\ea
where ${\bf S}\cdot{\bf S}=S(S+1)$ and the sum over $\langle ij\rangle$
is over nearest neighbors, counting each bond once. $J$ is the
exchange constant and ${\bf B}$ is the external magnetic field.
Positive anisotropy parameters
$K_2, K_4^\perp>0$ favor a perpendicular magnetization (easy-axis magnet),
while the opposite sign favors an in-plane magnetization (easy-plane magnet).
Two important special cases are $K_4^\|=0$, corresponding to uniaxial
symmetry, and $K_2=0$, $K_4^\|=K_4^\perp$, corresponding to cubic symmetry.
Equation (\ref{2.H1}) is viewed as a {\it microscopic\/} Hamiltonian for a
ferromagnetic monolayer, where the 
anisotropy parameters are obtained, for example, from first-principles
calculations. The same Hamiltonian can also be used approximately
as an {\it effective\/} model for films with
several layers, in which case the anisotropy constants are to be understood
as averaged parameters containing both bulk and surface contributions.
Since anisotropy energies (and also the exchange coupling $J$) can be quite
different for the surface layers and for the bulk, the effective parameters
can be tuned by changing the film thickness.

Equation (\ref{2.H1}) does not include the magnetic dipolar interaction
\be
H_{\text{dip}} = \frac{\omega}{2}\,\sum_{i,j\,(i\neq j)}
  \left[ \frac{{\bf S}_i\cdot{\bf S}_j}{r_{ij}^3}
  - \frac{3({\bf S}_i\cdot{\bf r}_{ij})({\bf S}_j\cdot{\bf r}_{ij})}
    {r_{ij}^5} \right] ,
\label{2.Hdip1}
\ee
where ${\bf r}_{ij}={\bf r}_i-{\bf r}_j$
is the separation vector of sites $i$ and $j$,
$\omega=\mu_0 g^2 \mu_B^2/4\pi$, $\mu_0$ is the permeability of free space,
$g$ is the Land\'e factor, and $\mu_B$ is the Bohr magneton.
The dipolar interaction is difficult to treat in the present
framework, mainly because of its long-range nature. One may incorporate
it in a simple mean-field way by replacing one of the
two spin operators in Eq.~(\ref{2.Hdip1})
by its thermal average, which is, in our notation,
the magnetization ${\bf M}$. In a continuum approximation this leads to
\be
H_{\text{dip}} \cong -\frac{1}{2}\,\frac{2\pi\omega}{3a^3} \,
  (M_x, M_y,-2M_z) \cdot \sum_i {\bf S}_i ,
\label{2.Hdip2}
\ee
where $a$ is the lattice constant. Equation (\ref{2.Hdip2})
describes the interaction with an effective {\it demagnetizing\/}
field. This term explicitly breaks spin symmetry in that it
prefers the magnetization to lie in the $xy$ plane, as is well known.
In this approximation, the rotational symmetry {\it within\/} the plane is
retained. A better description would also break this residual symmetry.
This is the reason why a magnetic monolayer has a finite in-plane
magnetization even if only the exchange coupling and the magnetic dipole
interaction are considered, as has been shown using a Green's-function
approach.\cite{Bru91,FJK99} Since we are mainly
interested in the effects of the single-ion anisotropy
terms in the Hamiltonian
and in spontaneous symmetry breaking, we
neglect $H_{\text{dip}}$ for most of this paper.

We now give an overview over the derivation:
After replacing the spin operators in the Hamiltonian (\ref{2.H1})
by bosons, the bosonic system is generalized from two to $N$
boson species. We mention that we take care to preserve the quantum
properties of the spin operators in the anisotropy terms
as we perform the limit $N\to\infty$.
Then the partition function $Z$ is written as a functional
integral over complex auxilliary fields. The constraint on boson number is
incorporated using a Lagrange-multiplier field and the exchange
interaction as well as the anisotropy terms are decoupled by means of
a Hubbard-Stratonovich transformation. Within the mean-field approximation
the auxilliary fields are then replaced by constants, which have to be
obtained from saddle-point equations.

The resulting mean-field free energy $F_0$ depends on the four auxilliary
parameters
$\overline{\bf P}$ and $\overline{\Lambda}$. One has to find the saddle
points of $F_0$ with respect to these parameters and check their
stability, as discussed in App.~\ref{app:a}. $\overline{\Lambda}$ can
be interpreted as a
chemical potential that globally enforces the constraint on boson number,
whereas the vector $\overline{\bf P}$ is related to the magnetization,
see below. Introducing rescaled anisotropy parameters\cite{newMillev}
\ba
\tilde K_2 & = & K_2\,\frac{S-1/2}{S} ,
\label{resK2} \\
\tilde K_4^{\perp,\|} & = & K_4^{\perp,\|} \,
  \frac{(S-1/2)(S-1)(S-3/2)}{S^3}
\label{resK4}
\ea
and the inverse temperature $\beta=1/k_BT$, the free energy reads
\ba
\beta F_0
  & \!=\! & - S\overline{\Lambda}
    + \frac{\beta \tilde K_2}{2}\, \overline{P}_z^2
  \nonumber \\
& & {}+ \frac{3\,\beta \tilde K_4^\perp}{2}\, \overline{P}_z^4
    + \frac{3\,\beta \tilde K_4^\|}{2}\,
    \left( \overline{P}_x^4 + \overline{P}_y^4 \right)
  \nonumber \\
& & {}+ \frac{1}{8\pi\beta JS}\, \Big[
      \text{diln} \left(1-e^{-\overline{\Lambda}+\beta B'/2}\right)
  \nonumber \\
& & \quad{}+ \text{diln} \left(1-e^{-\overline{\Lambda}-\beta B'/2}\right)
  \Big]
\label{2.F8}
\ea
with the dilogarithm function $\text{diln}$ and the effective magnetic
field
\ba
B'_x & = & B_x + 4\, \tilde K_4^\|\, \overline{P}_x^3 , \\
B'_y & = & B_y + 4\, \tilde K_4^\|\, \overline{P}_y^3 , \\
B'_z & = & B_z + 2\, \tilde K_2\, \overline{P}_z
   + 4\, \tilde K_4^\perp\, \overline{P}_z^3 .
\label{2.Bp8}
\ea
We emphasize that the whole mathematical apparatus of SU($N$) SBMF
theory is contained in Eqs.~(\ref{2.F8})--(\ref{2.M08}).
Note that these expressions reduce to the isotropic case\cite{RS,TGHS}
for $K_2=K_4^\perp=K_4^\|=0$ and that the second-order (fourth-order)
anisotropies only enter the results if $S\ge 1$ ($S\ge 2$).

The mean-field magnetization is given by
\ba
\lefteqn{
{\bf M}_0 \equiv -2\,\frac{d F_0}{d{\bf B}}
  = -\frac1{8\pi\beta JS}\,\frac{{\bf B}'}{B'}
  } \nonumber \\
& & \quad\times
  \left[ \ln\left(1-e^{-\overline{\Lambda}+\beta B'/2}\right)
    - \ln\left(1-e^{-\overline{\Lambda}-\beta B'/2}\right) \right] ,
\label{2.M08}
\ea
where $B'=|{\bf B}'|$.
As shown in App.~\ref{app:a}, $\overline{P}_i$ as obtained from the
saddle-point equations equals the magnetization component $M_{0,i}$,
{\it if\/} the free energy $F_0$ depends on $\overline{P}_i$.
On the other hand, if $F_0$ does not depend on $\overline{P}_i$,
the latter quantity can assume any value, while the magnetization
$M_{0,i}$ still has a well-defined value given by Eq.~(\ref{2.M08}).

Within the simplified treatment of the dipolar interaction discussed above,
a demagnetizing field proportional to
$(M_{0,x},M_{0,y},-2M_{0,z})$ would be added to ${\bf B}'$.
By using a Green's-function method it can be shown that this
yields an acceptable result for the magnetization
if the magnetic dipole coupling is small as
compared to the exchange coupling.\cite{Jenxx}

Finally, we give the results of O($N$) SBMF theory without
derivation. The free energy per boson reads
\ba
\lefteqn{
\beta F_0^{\text{O}(N)}
  = -\frac13 S\overline{\Lambda} + 3\,\beta \tilde K_2\,
    \overline{P}_z^2
  } \nonumber \\
& & \quad{}+ 81\,\beta \tilde K_4^\perp\, \overline{P}_z^4
    + 81\,\beta \tilde K_4^\|\,
    \left(\overline{P}_x^4+\overline{P}_y^4\right)
  \nonumber \\
& & \quad{}+ \frac1{12\pi\beta J S}
  \Big[
    \text{diln} \left(1-e^{-\overline{\Lambda}+\beta B'}\right)
  \nonumber \\
& & \qquad{}+ \text{diln} \left(1-e^{-\overline{\Lambda}}\right)
    + \text{diln} \left(1-e^{-\overline{\Lambda}-\beta B'}\right)
  \Big]
\ea
with the effective field
\ba
B'_x & = & B_x + 108\, \tilde K_4^\|\, \overline{P}_x^3 , \\
B'_y & = & B_y + 108\, \tilde K_4^\|\, \overline{P}_y^3 , \\
B'_z & = & B_z + 6\, \tilde K_2\, \overline{P}_z
      + 108\, \tilde K_4^\|\, \overline{P}_z^3 ,
\ea
and the magnetization is
\ba
\lefteqn{
{\bf M}_0^{\text{O}(N)} = -\frac1{4\pi\beta JS}\,\frac{{\bf B}'}{B'}
  } \nonumber \\
& & \quad\times
  \left[ \ln\left(1-e^{-\overline{\Lambda}+\beta B'}\right)
  - \ln\left(1-e^{-\overline{\Lambda}-\beta B'}\right) \right] .
\ea
The general form of these expressions is similar to the SU($N$) case
and we thus expect qualitatively similar results.

\section{Results and discussion}
\label{sec:res}

In this section we discuss a number of results obtained from
SBMF theory and compare them with other approaches. Our aim is mainly to
show that SBMF theory yields qualitatively correct results in the
presence of single-ion anisotropies. We further
show that these  results are comparable to Green's-function
methods, although they are much easier to obtain. We first discuss the
magnetization $M_0$ and then effective anisotropies.

\subsection{Magnetization}
\label{sus:1}

\begin{figure}[h]
\centerline{\epsfxsize 3.00in\epsfbox{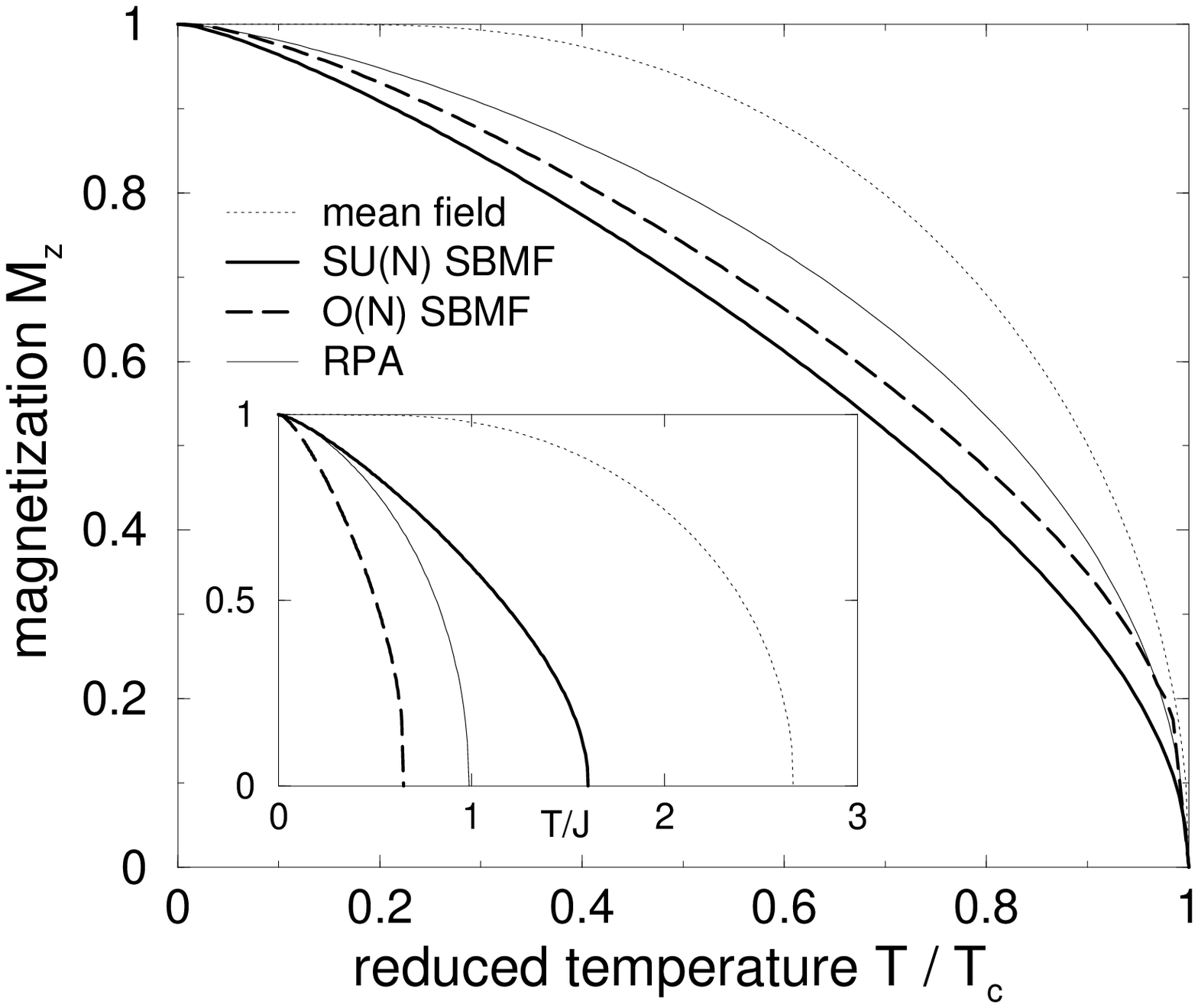}}
%\centerline{\epsfxsize 5.00in\epsfbox{anifig1.eps}}
\caption{Magnetization as predicted by various theories
as a function of reduced temperature $T/T_c$ for a film in a
vanishing external magnetic field, ${\bf B}=0$, with
an easy axis defined by $K_2/J=0.01$, spin $S=1$,
and no higher-order anisotropies. The magnetic dipole coupling
is neglected.
The inset shows the magnetization as a function of the absolute
temperature $T/J$ in units of exchange $J$. The line denoted by
`RPA' refers to a calculation within a many-body Green's-function
approach with Anderson-Callen
decoupling of the single-ion anisotropy.\protect\cite{FJK99}}
\label{fig1}
\end{figure}

SBMF theory correctly predicts the magnetization of an isotropic
2D Heisenberg ferromagnet to vanish at finite temperatures
in the absence of an external magnetic field.\cite{RS}
However, an easy-axis uniaxial anisotropy $K_2>0$
explicitly breaks the SU(2) spin symmetry down to a residual Ising symmetry.
A discrete symmetry can be spontaneously broken
and one expects a finite-temperature phase transition
between an ordered and a disordered phase. Both SU($N$)
and O($N$) theory indeed yield a finite $T_c$. Figure~\ref{fig1} shows
the magnetization of a spin-$1$ system
with a second-order uniaxial anisotropy $K_2/J=0.01$
in the absence of an external magnetic field,
as a function of the reduced temperature $T/T_c$. As usual,
simple mean-field theory overestimates
the magnetization. In particular, at low temperatures
the magnetization approaches ${\bf M}(T=0)$ exponentially
due to the neglect of spin waves. On the other hand,
SU($N$) and O($N$) Schwinger boson
theories and many-body Green's-function methods\cite{FJK99}
correctly predict an almost linear behavior. However, the
absolute temperature scales
are rather different, as shown in the inset of
Fig.~\ref{fig1}. Numerical results, {\it e.g.},
from Quantum-Monte-Carlo simulations, would be very useful to check the
absolute value of $T_c$.

One can obtain analytical expressions for $T_c$ by expanding
the saddle-point equations for a small magnetization $M_{0,z}$. The results
are\cite{rem:Qspin}
\be
T_c^{\text{SU}(N)} = \frac{4\pi JS^2}{\ln[1+4\pi JS^2/K_2(S-1/2)]}
\label{3.tc1}
\ee
for SU($N$) theory and
\be
T_c^{\text{O}(N)} = \frac{4\pi JS^2}{3\ln[1+\pi JS^2/K_2(S-1/2)]}
\label{3.tc2}
\ee
for O($N$) theory. Both expressions have the same functional form.
In particular, $T_c$ vanishes in the limit $K_2\to 0^+$, as it should,
and $T_c=0$ for $S=1/2$.
%%Note that $T_c$ depends on $K_2$ through
%%the first excitation energy $\Delta e=K_2 (2S-1)$
%%of the second-order anisotropy term $-K_2 (S^z)^2$
%%in units of the interaction
%%energy scale $JS^2$. This ratio is a measure of how strongly
%%the continuous SU(2) symmetry is broken.
For comparison, using a continuum approximation
similar to the one employed here, the many-body Green's-function
method yields
\be
%% check "additional" factor S above!
T_c^{\text{RPA}} = \frac{4\pi JS(S+1)}{3\ln[1+3\pi^2 JS/4K_2(S-1/2)]}
\label{3.tc3}
\ee
within the Anderson-Callen decoupling of the second-order
anisotropy.\cite{And64} Note that all three expressions show the
logarithmic dependence of $T_c$ on $K_2$, while the dependence
on the spin quantum number differs.

%%For large $K_2\gg T$, the anisotropic
%%Heisenberg model should approach the Ising model. In this limit the
%%critical temperature should become proportional to $JS^2$ and be
%%independent of $K_2$. However,
%%Eqs.~(\ref{3.tc1}) and (\ref{3.tc2}) yield
%%$T_c^{\text{SU}(N)}\cong K_2S$ and
%%$T_c^{\text{O}(N)}\cong 4K_2S/3$. Thus SBMF
%%theory does not give the correct behavior in the Ising limit. Note
%%that the Green's-function method suffers from the same problem.

Since the results of SU($N$) and O($N$) theory qualitatively agree
with each other, we concentrate on SU($N$) in the following.
In the case of a hard axis, $K_2<0$, and $K_4^\|=0$,
the residual symmetry is of $XY$ type.
Because of the continuous degeneracy there should be
no spontaneous magnetization.\cite{MW}
This is indeed found within SBMF theory.
If we include a dipolar interaction by means of Eq.~(\ref{2.Hdip2}),
a finite in-plane magnetization emerges, since in this approximation
the dipole coupling acts similarly to an external field.

\begin{figure}[h]
\centerline{\epsfxsize 3.00in\epsfbox{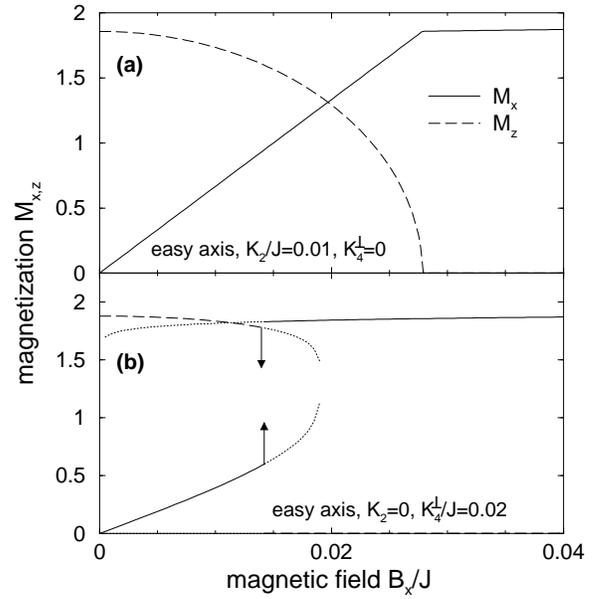}}
%\centerline{\epsfxsize 5.00in\epsfbox{anifig2.eps}}
\caption{Reorientation transitions as a function of external magnetic
field for a film with exchange $J=100$ and
spin $S=2$ at temperature $T/J=1$. (a) Film with second-order
easy-axis anisotropy $K_2/J=0.01$ and $K_4^\perp=K_4^\|=0$.
The field is applied in the $x$
direction, {\it i.e.}, perpendicularly to the easy axis. The solid
and dashed lines represent the $x$ and $z$ components of the magnetization.
In this case a second-order reorientation transition takes place.
(b) The same quantities
for a film with forth-order easy-axis anisotropy $K_4^\perp/J=0.02$
and $K_2=K_4^\|=0$. Here, a first-order reorientation transition takes
place, denoted by the arrows. The dotted lines refer to meta-stable states
of higher energy.}
\label{fig2}
\end{figure}

In the presence of an external magnetic field ${\bf B}$ one
always expects a finite magnetization. For an isotropic film this
magnetization is parallel to the applied field,
whereas in the presence of anisotropy the magnetization may be
oriented along a different direction. In addition, a
magnetic reorientation transition can occur, which may be controlled
by varying the temperature
$T$, the external magnetic field ${\bf B}$, or the film thickness.
In Fig.~\ref{fig2}(a) the magnetization components parallel and
perpendicular to the magnetic field are shown as functions
of its strength. The field is applied perpendicularly to the easy axis,
which is realized by a second-order uniaxial anisotropy $K_2/J=0.01$.
Within our theory we find a linear increase of the in-plane
magnetization along the magnetic-field direction below a critical
reorientation field $B_{\text{reo}}$. Since the modulus of the
magnetization is almost constant, the perpendicular magnetization
component decreases correspondingly. At
$B=B_{\text{reo}}$ the latter component vanishes continuously,
corresponding to a second-order phase transition. For
$B\ge B_{\text{reo}}$ the magnetization is parallel to the field and
increases only weakly with $B$. Figure~\ref{fig2}(b) shows a similar
situation,
with the easy axis realized by the {\it fourth\/}-order uniaxial anisotropy
$K_4^\perp>0$, while $K_2=K_4^\|=0$. Again a reorientation
transition is obtained as a function of the in-plane magnetic field.
However, since here the magnetization changes
discontinuously, the transition is of first order.
We expect that the dependence of the order of the transition on the
order of the uniaxial anisotropy is a universal feature, since the same
behavior has been observed within simpler
theories.\cite{Jen90,MOK}

\begin{figure}[h]
\centerline{\epsfxsize 3.00in\epsfbox{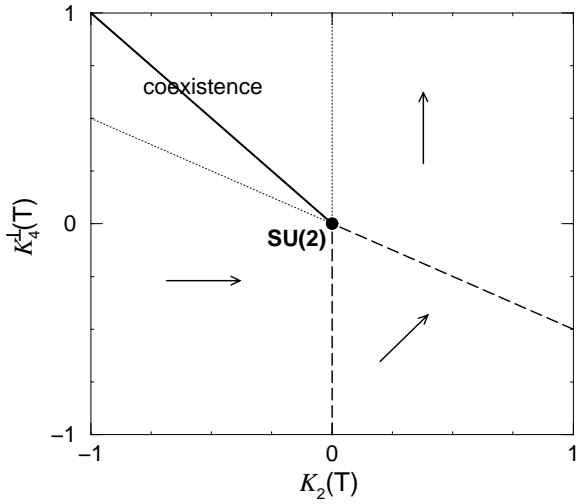}}
%\centerline{\epsfxsize 5.00in\epsfbox{anifig3.eps}}
\caption{Phase diagram for the {\it effective\/} anisotropies
${\cal K}_2(T)$ and ${\cal K}_4^\perp(T)$ for a vanishing
external magnetic field und ${\cal K}_4^\|=0$.\protect\cite{Cas59,Bab97,MOK}
Three different phases are found, defined by different directions of the
magnetization with respect to the film plane (perpendicular, in-plane,
canted), indicated by arrows. These phases are separated either by
a first order (solid line) or by a second order phase transition
(dashed lines), which merge at the tricritical point
${\cal K}_{2c}={\cal K}_{4c}^\perp=0$, corresponding to
the isotropic, SU(2)-symmetric case. A coexistence region
between the perpendicular and in-plane phases
exists near the first order phase transition boundary, indicated by the
dotted lines.}
\label{fig3}
\end{figure}

To show the interplay of the various interaction terms, we
now discuss several phase diagrams. For comparison, Fig.~\ref{fig3} shows
the well-known phase diagram in the ${\cal K}_2$-${\cal K}_4^\perp$
plane for ${\cal K}_4^\|=0$ and ${\bf B}=0$ derived under the assumption
that the magnetization behaves like a macroscopic, classical magnetic
moment.\cite{Cas59,Bab97,MOK}
The anisotropies should be viewed as {\it effective\/} quantities, which
depend on temperature etc. Roughly,
a perpendicular magnetization is found for ${\cal K}_2>0$ and
${\cal K}_4^\perp>0$,\cite{rem:domains} an
in-plane one for ${\cal K}_2<0$ and ${\cal K}_4^\perp<0$, and a canted
one for ${\cal K}_2>0$ and ${\cal K}_4^\perp<0$. These magnetic phases
are separated by
continuous (second order) or discontinuous (first order) phase transition
boundaries, which merge in a tricritical point. A
coexistence region between the perpendicular and the in-plane magnetization
is located close to the first-order phase transition line.
As usual, the first-order phase boundary has
been drawn where the free energies of both phases are equal.

The simple phase diagram of Fig.~\ref{fig3} does not apply to 2D systems,
since there should be no net in-plane magnetization
if solely uniaxial anisotropies are taken into
account. The reason is that in this case a ground state with finite in-plane
magnetization would be continuously
degenerate.\cite{MW} The origin of this problem of course lies in the
neglect of fluctuations.
%%To incorporate their effect in a crude way, we may replace in
%%Fig.~\ref{fig3} the in-plane magnetized phase by one with vanishing
%%magnetization, and retain for the canted phase
%%only the perpendicular magnetization component. In doing so, the difference
%%between the perpendicular and the canted phases vanishes, and the
%%second-order transition for ${\cal K}_2>0$ in Fig.~\ref{fig3}
%%disappears.\cite{rem:domains}

\begin{figure}[h]
\centerline{\epsfxsize 3.00in\epsfbox{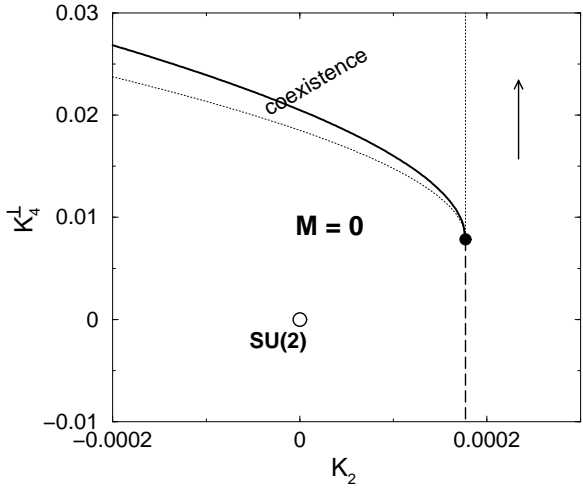}}
%\centerline{\epsfxsize 5.00in\epsfbox{anifig4.eps}}
\caption{Phase diagram of a thin film in the $K_2$-$K_4^\perp$ plane
in a vanishing external magnetic field as calculated
from SBMF theory. The parameters are
spin $S=2$, exchange $J=100$, temperature $T/J=3$, and in-plane
anisotropy $K_4^\|=0$.
The phase to the lower left of the phase diagram now exhibits no
net magnetization,
${\bf M}=0$. The open circle denotes the isotropic,
SU(2)-symmetric case $K_2=K_4^\perp=0$.
The other symbols are the same as in Fig.~\protect\ref{fig3}.}
\label{fig4}
\end{figure}

This picture is changed if fluctuations are included by means of
SU($N$) SBMF theory:
Figure \ref{fig4} shows the phase diagram in the $K_2$-$K_4^\perp$
plane for a thin film at a finite temperature $T/J=3$ for
spin $S=2$.\cite{rem:spin}
The anisotropy parameters are now the microscopic ones contained in
the Hamiltonian (\ref{2.H1}).
Note that the scale of the anisotropies in Fig.~\ref{fig4}
is small, since the temperature $T/J=3$ is significantly lower than the
typical temperature scale $4\pi J S^2$ of Eq.~(\ref{3.tc1}). This
means that very small anisotropies are sufficient to induce a finite
magnetization, since $K_2$ enters logarithmically in $T_c$,
Eq.~(\ref{3.tc1}).

%%The topology of the phase diagram in Fig.~\ref{fig4} is the same as
%%inferred from Fig.~\ref{fig3} by including the action of
%%fluctuations by hand.
In Fig.~\ref{fig4} only two phases are present: A phase without net
magnetization located approximately in the region $K_2<0$ and $K_4^\perp<0$
and a phase with perpendicular magnetization otherwise. These
two phases are separated by either a first-order or a second-order
transition. The tricritical point, at which the two
merge, has shifted to finite, temperature-dependend
values of the anisotropies $K_{2c}>0$ and $K_{4c}^\perp>0$.
For smaller anisotropies thermal
fluctuations overcome the tendency to order.
A coexistence region is found in the vicinity
of the first-order transition. The first-order
line is no longer straight
and approaches the tricritical point with a vertical tangent. For negative
$K_4^\perp$ the second-order transition line is given by
$K_2=\text{const}$, similar to Fig.~\ref{fig3}.
Thus the location of this transition depends only on $K_2$.
At this transition the zero-magnetization (${\bf M}=0$) saddle
point becomes unstable. The free energy for small ${\bf M}$,
and thus the stability of this saddle point, is solely determined by
the lowest-order anisotropy $K_2$.

\begin{figure}[h]
\centerline{\epsfxsize 3.00in\epsfbox{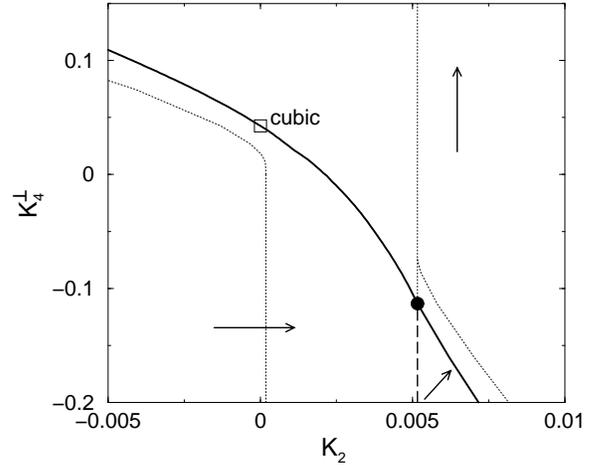}}
%\centerline{\epsfxsize 5.00in\epsfbox{anifig5.eps}}
\caption{Phase diagram for a thin film as in Fig.~\protect\ref{fig4}
but with a finite in-plane magnetic anisotropy
$K_4^\|/J\approx 4.3\times 10^{-4}$, big enough to allow a finite
in-plane magnetization
at this temperature. The square denotes the point of
full cubic symmetry ($K_2=0$, $K_4^\perp=K_4^\|$). The dotted lines are
borders of two coexistence regions. In the left one both perpendicular and
in-plane magnetization are metastable, while in the right one perpendicular
and canted states are metastable.}
\label{fig5}
\end{figure}

In-plane and canted magnetic phases do not appear in Fig.~\ref{fig4},
since this
system is either in the Ising or in the $XY$ universality class.
However, such phases do exist if the in-plane symmetry is explicitely
broken, {\it e.g.}, by a single-ion in-plane anisotropy
$K_4^\|\neq 0$, by an external magnetic field, or by the dipole
interaction. Here we consider a small fourth-order
in-plane anisotropy
$K_4^\|/J\approx 4.3\times 10^{-4}>0$. The other parameters are unchanged.
The resulting phase diagram is shown in
Fig.~\ref{fig5}, exhibiting now three phases similar to
Fig.~\ref{fig3}. However,
the transition between the canted and perpendicular phases
is now of first order (the first-order line extends to
arbitrarily large negative $K_4^\perp$). The first-order character
is typically weak---the jump in the magnetization is rather small.
The coexistence regions are also shown in Fig.~\ref{fig5}.
Note that for even smaller $K_4^\|$ thermal fluctuations would destroy
the in-plane magnetization, resulting in a phase diagram similar to
Fig.~\ref{fig4}.

Considering now a thin film with a finite number $n$ of atomic layers,
the anisotropies have to be viewed as averages over all layers and
can be tuned by varying $n$.
A variation of the film thickness thus refers
to a certain trajectory (``anisotropy flow''\cite{MOK})
in the corresponding phase diagram, {\it e.g.}, Fig.~\ref{fig5}.
In this way we can in principle describe reorientation transitions
as a function of film thickness.

\begin{figure}[h]
\centerline{\epsfxsize 3.00in\epsfbox{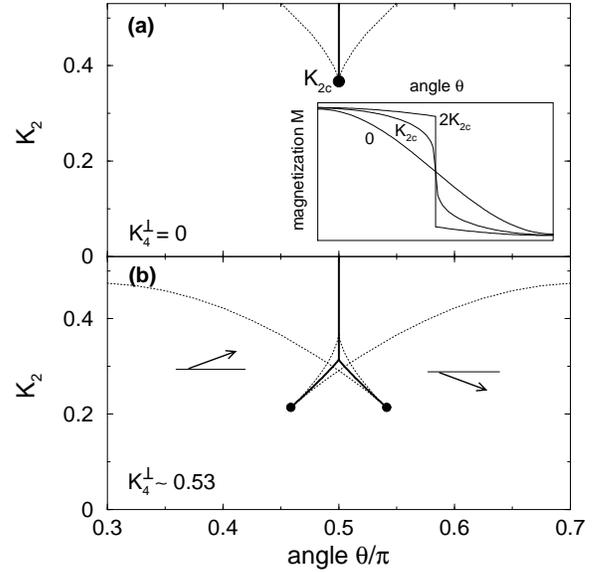}}
%\centerline{\epsfxsize 5.00in\epsfbox{anifig6.eps}}
\caption{Phase diagram for a thin film in the $K_2$-$\theta$ plane
as calculated from SBMF theory. An easy-axis
magnet is assumed, with $K_2>0$.
A magnetic field of strength $B/J=0.01$ is applied
in the $xz$ plane, forming an angle $\theta$ with the film normal.
We assume $J=100$, $S=2$, $T=J$, and $K_4^\|=0$. Furthermore, in (a)
$K_4^\perp=0$, whereas in (b) $K_4^\perp/J\approx 5.3 \times 10^{-3}$.
The thick solid lines denote first-order transitions ending in
critical points. The dotted lines are the boundaries of coexistence
regions. The inset shows the magnetization as a function of $\theta$
for the case (a) with $K_2=0$, $K_2=K_{2c}$,
and $K_2=2K_{2c}$, $K_{2c}$.}
\label{fig6}
\end{figure}

Finally, we consider the effect of a finite magnetic field. We keep
the absolute value of the field constant, $B/J=0.01$, and
change only its direction. For simplicity we restrict ourselves to
the case $K_4^\|=0$, and consider the case $S=2$ and $T/J=1$ as an example.
The magnetic field is rotated in the $xz$ plane from $\theta=0$
to $\theta=\pi$, where $\theta$ is the angle between the field and
the film normal. Figure~\ref{fig6}(a) shows the phase diagram for
the case $K_4^\perp=0$.
The magnetization tries to be aligned parallel to the magnetic field,
but is hindered by the anisotropies. For a sufficiently small
$K_2$ the magnetization follows the field continuously, whereas
for $K_2$ larger than a critical value $K_{2c}$ it exhibits a jump.
This behavior is illustrated by the inset in Fig.~\ref{fig6}(a), which
shows the perpendicular magnetization component as a function of
$\theta$ for $K_2=0$, $K_2=K_{2c}$, and $K_2=2K_{2c}$.
A negative value of $K_4^\perp$ does not change this behavior,
except that the coexistence regions becomes more narrow.
For $K_4^\perp>0$, however, the first-order line extends to smaller
$K_2$ and splits into a two-prong fork, as shown in Fig.~\ref{fig6}(b).
Each prong ends in a critical point. The metastability regions are
also indicated in Fig.~\ref{fig6}(b). The fork-like feature grows for
increasing $K_4^\perp$ and the critical points even enter the negative
$K_2$ region for sufficiently large $K_4^\perp$.
As a side remark, in the language of catastrophy theory the phase
diagram in the $(K_2, K_4^\perp, \theta)$ space
shown in Fig.~\ref{fig6} can be described by a {\it butterfly\/}
catastrophy. The special cut $K_4^\perp=0$, Fig.~\ref{fig6}(a),
shows a {\it cusp}.\cite{cata}

\subsection{Effective anisotropies}
\label{sus:2}

The effective magnetic anisotropies are most generally defined through
the dependence of the free energy on the magnetization direction.
The anisotropic part of the free energy
$F(T,\theta,\phi)$ ($\theta$ and $\phi$ being the polar and azimutal angles
of the magnetization)
depends on the symmetry of the lattice and is usually written with the
help of spherical harmonics Y$_l^m$,\cite{Cal66,Mil95} or as a series in
powers of the direction cosines.\cite{JeB98} For
example, we give the expression for
tetragonal symmetry, including also the Zeeman term,
\ba
F & = & -{\bf B}\cdot{\bf M} - {\cal K}_2 \cos^2\theta
  -{\cal K}_3 \cos^3\theta
  -{\cal K}_4^\perp \cos^4\theta
  \nonumber \\
& & {}-{\cal K}_4^\| \sin^4\theta\, (3/4+\cos 4\phi)
  -\ldots
\label{5.Eef1}
\ea
The {\it effective\/} anisotropies ${\cal K}_2(T)$ etc.\
depend strongly on temperature. They have to be
distinguished from the {\it microscopic\/} anisotropies $K_2$ etc.
appearing in the Hamiltonian (\ref{2.H1}).
Experimentally, the effective anisotropies are determined
from, {\it e.g.}, ferromagnetic resonance (FMR) experiments.
To compare the microscopic anisotropies calculated at $T=0$
with measurements of the effective anisotropies performed at
finite temperatures, the temperature dependence of the latter has to
be known. This task will be pursued in the
present section by use of the Schwinger boson technique.

The main source of the temperature dependence of the effective anisotropies
is the decreasing magnetization {\bf M}$(T)$.\cite{JeB98} Other sources,
such as the population change of spin-orbit splitted energy levels near the
Fermi energy,\cite{MHB96} will not be considered here, since the thermal
variation due to these effects occurs mainly at higher temperatures.
Note that the effective anisotropies depend on microscopic anisotropy terms
of various orders,\cite{Mos94} and can
even be present at finite temperatures if corresponding microscopic
ones are absent from the Hamiltonian.\cite{JeB98}

%%One way to describe the magnetic response of a thin film is through
%%the use of effective anisotropies.
%%In FMR experiments the magnetization typically behaves like
%%a single, macroscopic magnetic moment, which can be
%%perturbed out of equilibrium as a whole and then precesses
%%around the magnetic field direction. The precession frequency
%%provides information on the effective anisotropies.\cite{FMR,Bab97}
%%The magnetization acts as a classical moment if
%%spatial fluctuations are weak, which is the case for large
%%exchange coupling. However, the absolute value $M$ of the
%%magnetization is reduced for less favorable orientations.
%%Formally, this follows from the reduced mean field ${\bf B}'$.
%%Thus, $M$ will be a non-trivial function of the orientation
%%$\hat{\bf m}$. This mechanism can generate higher-order effective
%%anisotropies even if the corresponding microscopic term is absent.

For the determination of the effective anisotropies we proceed similarly
as in Ref.~\onlinecite{JeB98}. In order to turn the magnetization
into a direction $(\theta,\phi)$
different from the equilibrium one, we apply
an auxiliary magnetic field ${\bf B}$.
For the magnetization we use the SBMF result ${\bf M}_0$,
Eq.~(\ref{2.M08}). There are infinitely many choices of the magnetic
field ${\bf B}$ that result in the same magnetization direction. This
freedom of choice introduces some arbitrariness,
since the resulting effective anisotropies depend on the choice of ${\bf
B}$. We here choose the field ${\bf B}=(B,0,0)$ along the $x$ axis and vary
only its strength.\cite{JeB98} We insert the SBMF result for the free energy
$F_0$, Eq.~(\ref{2.F8}), into Eq.~(\ref{5.Eef1}). To obtain the effective
anisotropies we calculate the angles $\theta$, $\phi$ and the free energy
parametrically for a sufficient number of values of the magnetic-field
strength $B$ and fit Eq.~(\ref{5.Eef1}) to the resulting data.

\begin{figure}[h]
\centerline{\epsfxsize 3.00in\epsfbox{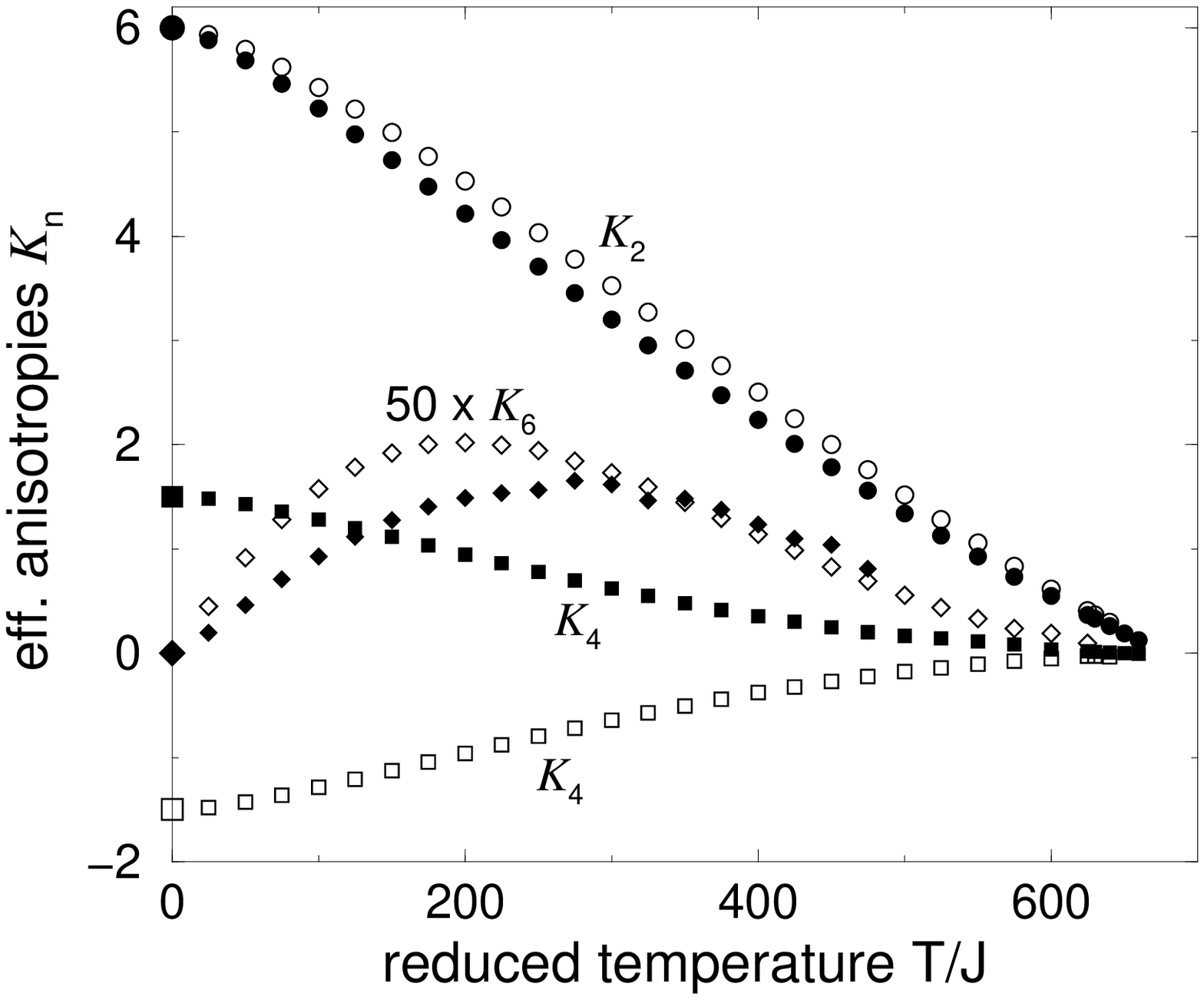}}
%\centerline{\epsfxsize 5.00in\epsfbox{anifig7.eps}}
\caption{Effective anisotropies ${\cal K}_2$, ${\cal K}_4$,
and ${\cal K}_6$ as functions of temperature calculated from
SBMF theory. A thin film is assumed with
$S=2$, $J=100$, $K_2/J=0.01$, $K_4^\|=0$, and $K_4^\perp/J=10^{-3}$
(filled symbols) and $K_4^\perp/J=-10^{-3}$ (open symbols).
We use rather large microscopic anisotropies to demonstrate more clearly
the appearance of a small sixth-order effective anisotropy ${\cal K}_6$,
which is shown scaled by a factor of 50. The big symbols denote the exact
values at $T=0$.}
\label{fig7}
\end{figure}

We now apply this method to a system with uniaxial easy-axis anisotropy.
In this case the free energy does not depend on the angle $\phi$ in
the absence of a magnetic field and hence the effective anisotropies
involving $\phi$ vanish. Then the free energy has the form
\be
F_0 + {\bf B}\cdot{\bf M} =  - \sum_{n=2}^\infty {\cal K}_n
  \cos^n\theta .
\label{5.Eef2}
\ee
We calculate the free energy $F_0$ for equally spaced values of the
direction cosine $\cos\theta$ by varying the field strength $B$. If a
first-order reorientation transition takes place at some field value
$B_{\text{reo}}$ we use only fields $B\le B_{\text{reo}}$.
We finally obtain the effective anisotropies by a least-square fit.
Figure~\ref{fig7} shows the effective anisotropies for
a thin film with $S=2$, $J=100$, and
uniaxial anisotropies $K_2/J=0.02$,
$K_4^\perp/J=0.01$ (filled symbols) and $K_2/J=0.02$,
$K_4^\perp/J=-0.01$ (open symbols). We here consider relatively large
anisotropies for illustrative purposes
so that higher-order anisotropies do not vanish
in the numerical errors. The odd effective anisotropies
${\cal K}_3$ etc.\ vanish because of symmetry. A small
sixth-order effective anisotropy ${\cal K}_6(T)$ appears at finite
temperatures, as well as anisotropies of even higher order, which are
much smaller, although no corresponding terms are present in the
Hamiltonian. The exact values of ${\cal K}_{2,4}(0)$ for $T\to 0$ are
${\cal K}_2(0)=K_2\,S(S-1/2)=\tilde K_2 S^2$ and
${\cal K}_4(0)=K_4^\perp\,S(S-1/2)(S-1)(S-3/2)=
  \tilde K_4^\perp S^4$,\cite{Cal66,Mil95,JeB98} cf.\ Eqs.~(\ref{resK2})
and (\ref{resK4}).
Our approach reproduces these limiting values. Note that the special
generalization of the anisotropy terms from SU(2) to SU($N$) discussed
in App.~\ref{app:a} is required to obtain these results. For $T\to T_c$
the effective anisotropies
vanish as ${\cal K}_n(T)\propto M^n(T)$.\cite{Cal66,Jen93} Therefore,
${\cal K}_4(T)$ decreases faster with increasing temperature than
${\cal K}_2(T)$ and exhibits a more smooth behavior near $T_c$.
The effective anisotropies shown in Fig.~\ref{fig7} qualitatively
agree with many-body Green's-function results.\cite{EFJK}

Finally, we emphasize that the calculation of effective
anisotropies with this method suffers from ambiguities due
to the freedom of choice of the magnetic field
and the particular fitting procedure employed.
Furthermore, this derivation employs a static
calculation of the magnetization and free energy in order to describe a
dynamical (FMR) experiment. These problems are common to all calculations
of effective anisotropies we are aware of. A full quantum-mechanical,
dynamical theory of FMR would be very useful. One possible approach
would be to incorporate spin dynamics into the Schwinger boson
theory.

\section{Conclusions and outlook}
\label{sec:fin}

In summary, we have presented the Schwinger boson mean-field
theory for anisotropic 2D ferromagnets, using two
different bosonizations.
In this study we have focused on the theoretical treatment of
single-ion anisotropies of a Heisenberg-type Hamiltonian.
The anisotropy energies break the full SU(2)
spin symmetry, leading to a lower residual symmetry, {\it e.g.},
of Ising or $XY$ type. We have shown that the results for the
magnetization satisfy the Mermin-Wagner theorem\cite{MW} also in these
cases, in that a finite spontaneous magnetization is only found
if the ground state is not continuously degenerate. The
theory can describe the various magnetic order-disorder transitions
as well as the reorientation transitions found in experiment.
The bosonization method can serve as an alternative technique
for quantum spin systems, besides the many-body Green's-function
approach.\cite{FJK99}
Our results for magnetization and effective anisotropies
qualitatively agree with the ones obtained by the latter method.
Microscopic calculations of effective anisotropies as functions
of temperature are important, since they---and not the microscopic
anisotropies appearing in the Hamiltonian---are obtained from
experiments. However, the absolute value of the critical temperature at
which the spontaneous magnetization becomes finite
strongly depends on the details of the theory.
Compared to the Green's-function method, Schwinger boson theory
has the advantage of treating any spin value $S$ and any
orientation of the magnetization relative to the anisotropy axes
and the external magnetic field on the same footing. It is also
numerically much less demanding.
Other types of anisotropies, for example of hexagonal
surfaces, can be incorporated in a straightforward manner.
Finally, Schwinger boson mean-field theory is a well-controlled
approximation in the sense that
fluctuations about the equilibrium solution can be
treated systematically within a $1/N$ expansion.\cite{Auerbach}

We conclude with enumerating possible generalizations of the theory
that are of particular interest for a realistic description of
magnetic thin films: The first is the explicit description of
thin films of several layers, where the exchange couplings,
the magnetic moments, and the
anisotropies are microscopic quantities which
depend on the layer index. This can be done by a straightforward
generalization of the theory, where one only has to keep track
of additional layer indices. The second is the inclusion of
the long-range magnetic dipole interaction. This
interaction is an additional source of anisotropy, which acts
quite differently from the single-ion anisotropies, resulting
for instance in the formation of magnetic domains.\cite{JeB95}
We reiterate that both the effect of
several layers and the dipole interaction can be treated in the
present theory as it is, albeit only in a mean-field manner.
The third interesting generalization is the inclusion of magnetic
dynamics. This would allow to describe the precession of
a non-equilibrium magnetization and constitute a microscopic theory of
ferromagnetic resonance. We leave these questions as projects for
the future.

\acknowledgements

We would like to thank K.H. Bennemann for helpful remarks.
We gratefully acknowledge financial support by the Deutsche
Forschungsgemeinschaft through Son\-der\-for\-schungs\-be\-reich 290,
TP A1.

\appendix

\section{}
\label{app:a}

In this appendix we present the Schwinger boson theory for anisotropic
ferromagnetic films in some more detail.
To derive the mean-field equations we start from the Hamiltonian
(\ref{2.H1}).
We replace the spin operators at each lattice site by two
Bose fields $b_{\up i}$ and $b_{\down i}$ according to
Eqs.~(\ref{1.S1})--(\ref{1.S3}). The fields $b_{\up i}$ and
$b_{\down i}$ have to satisfy the constraint (\ref{1.Sc1}) at each
lattice site. With the matrix
\be
\text{\sf S}(i) \equiv \left( \begin{array}{cc}
            b_{\up i}^\dagger b_{\up i} & b_{\up i}^\dagger b_{\down i} \\
            b_{\down i}^\dagger b_{\up i} & b_{\down i}^\dagger b_{\down i}
         \end{array} \right) ,
\ee
the Hamiltonian reads, up to an additive constant,
\ba
H & = & -\frac{J}2 \sum_{\langle ij\rangle} S_\beta^\alpha(i)
    S_\alpha^\beta(j)
  - \frac{\bf B}2 \sum_i \text{\boldmath$\sigma$}_\beta^\alpha
    S_\alpha^\beta(i)
  \nonumber \\
& & {}- \frac{K_2}4 \sum_i \left[ (\sigma^z)_\beta^\alpha S_\alpha^\beta(i)
    \right]^2
  - \frac{K_4^\perp}8 \sum_i \left[ (\sigma^z)_\beta^\alpha S_\alpha^\beta(i)
    \right]^4
  \nonumber \\
& & {}- \frac{K_4^\|}8 \sum_i \left(\left[
    (\sigma^x)_\beta^\alpha S_\alpha^\beta(i)\right]^4
    + \left[(\sigma^y)_\beta^\alpha S_\alpha^\beta(i)\right]^4 \right)
\ea
with the Pauli matrices
$\text{\boldmath$\sigma$}=(\sigma^x,\sigma^y,\sigma^z)$.
Summation over repeated greek indices is implied.

This expression is now generalized to $N$ Bose fields $b_\alpha$.
Then $\text{\sf S}$ becomes an $N\times N$ matrix of operators
$b_\alpha^\dagger b_\beta$ and the constraint is generalized to
\be
b_\alpha^\dagger b_\alpha=NS .
\label{2.bc1}
\ee
The Pauli matrices are generalized
to $N\times N$ block-diagonal matrices with the original Pauli
matrices repeated along the diagonal. We keep the notation $\sigma^z$ etc.

There is some freedom in the generalization of the Hamiltonian to $N$ boson
flavors---we mainly have to make sure that the correct $N=2$ case is
recovered. If we naively generalize the second-order anisotropy term as
$-(K_2/2N)\,[\sum_{\alpha\beta} (\sigma^z)^\alpha_\beta S^\beta_\alpha]^2$,
we run into problems, which are most easily illustrated for the case
$S=1/2$. For $N=2$ the operator $S^z$ has the eigenvalues $\pm 1/2$ so that
$(S^z)^2$ only has the eigenvalue $1/4$. This is, of course, the reason why
the $K_2$ term is irrelevant for $S=1/2$. However, for $S=1/2$ and
$N>2$ ($N$ even) we find $N/2+1$ different eigenvalues of the $z$ spin
component $(\sigma^z)^\alpha_\beta S^\beta_\alpha$ and $N/4+1$ (or $N/4+1/2$)
different eigenvalues of the $K_2$ term for $N$ (not) an integer multiple
of $4$. Thus in the limit $N\to\infty$ the second-order anisotropy term has
infinitely many different eigenvalues even for $S=1/2$, which looks like a
semi-classical approximation. Clearly, the mean-field results would depend
on $K_2$ for $S=1/2$, which is incorrect. To remedy this problem we
introduce an $N$-dependent anisotropy $K_{2,N}$ such that the energy
difference between the two lowest energy states of the second-order
anisotropy term,
\be
\Delta e = \frac{NK_{2,N}}{2}
  \left[S^2 - (S-2/N)^2\right] = 2K_{2,N} S - \frac{2K_{2,N}}{N}
\ee
is kept constant for all $N$. In
particular, $\Delta e=0$ for $S=1/2$ and all $N$. This is achieved by
setting
\be
K_{2,N} = \frac{S-1/2}{S-1/N}\,K_2 .
\label{2.K2N}
\ee
In the mean-field approximation, $N\to\infty$, we get\cite{newMillev}
\be
\tilde K_2\equiv K_{2,\infty}=K_2\,(S-1/2)/S .
\ee
Analogously, the $N$-dependent fourth-order
anisotropies are obtained,
\be
K_{4,N}^{\perp,\|} = \frac{(S-1/2)(S-1)(S-3/2)}{(S-1/N)(S-2/N)(S-3/N)}\,
  K_4^{\perp,\|}
\label{2.K4N}
\ee
and\cite{newMillev}
\be
\tilde K_4^{\perp,\|} \equiv K_{4,\infty}^{\perp,\|}
  = K_4^{\perp,\|}\,(S-1/2)(S-1)(S-3/2)/S^3 .
\ee
The fourth-order anisotropy contributions now correctly vanish for $S<2$
even at the mean-field level.

In order to write down a functional integral over the Bose fields,
we have to normal order the bosonic operators. The exchange term is
trivial, since the operators at different lattices sites commute.
By commuting the operators in the anisotropy terms, using the constraint
(\ref{2.bc1}), and dropping a constant we obtain
\ba
\lefteqn{
H_{\text{SU}(N)} = -\frac{J}{N} \sum_{\langle ij\rangle}
    S_\beta^\alpha(i) S_\alpha^\beta(j)
  - \frac{\bf B}2 \sum_i \text{\boldmath$\sigma$}_\beta^\alpha
    S_\alpha^\beta(i)
  } \nonumber \\
& & \quad\!{}- \frac{K'_{2,N}}{2N} \sum_i :\!\left[ (\sigma^z)_\beta^\alpha
    S_\alpha^\beta(i) \right]^2\!:
  \nonumber \\
& & \quad\!{}-
  \frac{K_{4,N}^\perp}{2N^3} \sum_i :\!\left[ (\sigma^z)_\beta^\alpha
    S_\alpha^\beta(i) \right]^4\!:
  \nonumber \\
& & \quad\!{}- \frac{K_{4,N}^\|}{2N^3} \sum_i :\!\left(\left[
    (\sigma^x)_\beta^\alpha S_\alpha^\beta(i)\right]^4
    + \left[(\sigma^y)_\beta^\alpha S_\alpha^\beta(i)\right]^4 \right)\!: ,
  \nonumber \\
& & {}
\ea
where $:\::$ denote normal ordering. The second-order anisotropy has
obtained a contribution from commuting the bosons in the fourth-order
terms, similar to other approaches,\cite{Jen93,JeB98}
\be
K'_{2,N} = K_{2,N} + \frac{6NS+4}{N^2}\,K_{4,N}^\perp -
  \frac{6NS+4}{N^2}\,K_{4,N}^\| .
\ee
If we want to do a consistent theory to order zero in $1/N$ we can just
drop the corrections. However, since the prefactors
$(6NS+4)/N^2 = 3S+1$ are not small for $N=2$ it may be better to keep
them. Note also that they cancel for cubic symmetry. In the following
we omit the prime at $K_2$ but keep these points in mind.

Next, we go over to the continuum limit\cite{TGHS} and write down
the functional integral for the partition function. The constraint
(\ref{2.bc1}) is implemented using a Lagrange multiplier
$\lambda({\bf r},\tau)$ at each point in space ${\bf r}$ and imaginary
time $\tau$.\cite{Auerbach} $\lambda$ is to be integrated from $-i\infty$
to $+i\infty$ but we can deform the path of integration between these
limits. The path of integration has to cross the real axis on the positive
side to avoid a branch cut. The partition function reads
\be
Z = \int D^2b_\alpha\,D\lambda\,
  \exp\!\left( -\frac1{\hbar}
  \int_0^{\hbar\beta}
  \!\!\!d\tau \int d^2r\, {\cal L}[b;\lambda] \right) ,
\label{2.Zf1}
\ee
where now $b_\alpha({\bf r},\tau)$ is a complex field. The Lagrangian is
\ba
\lefteqn{
{\cal L}[b;\lambda] = \frac{\hbar}{a^2} b_\alpha^\ast
    \partial_\tau b_\alpha
  + JS (\partial_j b_\alpha^\ast)(\partial_j b_\alpha)
  } \nonumber \\
& & \quad{}- \frac{J}{N}\, b_\alpha^\ast(\partial_j b_\beta^\ast)
    b_\beta(\partial_j b_\alpha)
  - \frac{\bf B}{2a^2}\, \text{\boldmath$\sigma$}_\beta^\alpha
    b_\beta^\ast b_\alpha
  \nonumber \\
& & \quad{}- \frac{K_{2,N}}{2N a^2}\, \left[ (\sigma^z)_\beta^\alpha b_\beta^\ast
    b_\alpha \right]^2
  - \frac{K_{4,N}^\perp}{2N^3 a^2}\, \left[ (\sigma^z)_\beta^\alpha b_\beta^\ast
    b_\alpha \right]^4
  \nonumber \\
& & \quad{}- \frac{K_{4,N}^\|}{2N^3 a^2}\,\left( \left[(\sigma^x)_\beta^\alpha
    b_\beta^\ast
    b_\alpha \right]^4 + \left[(\sigma^y)_\beta^\alpha b_\beta^\ast
    b_\alpha \right]^4 \right)
  \nonumber \\
& & {}+ \lambda b_\alpha^\ast b_\alpha - NS \lambda .
\ea
Here, $a$ is the lattice spacing, $\partial_\tau$ is the time derivative,
and $\partial_j$ is a spatial derivative along the $j$ direction.
Summation over repeated indices is implied.

In the next step, the terms containing more than two $b_\alpha$ fields
are decoupled using Hubbard-Stratonovich transformations,
thereby introducing additional auxiliary fields ${\bf Q}$, ${\bf P}$, and
${\bf R}$, which have to be integrated over in the functional integral.
In a short-hand notation the decoupling of the exchange term
may be written\cite{TGHS}
\ba
\lefteqn{
- \frac{J}{N}\, b_\alpha^\ast(\partial_j b_\beta^\ast)
    b_\beta(\partial_j b_\alpha)
  \to NJ Q_jQ_j
  + iJ Q_j\, b_\alpha^\ast(\partial_j b_\alpha)
  } \nonumber \\
& & \qquad{}- iJ Q_j\, (\partial_j b_\alpha^\ast) b_\alpha .
  \qquad\qquad\qquad\qquad\qquad
\ea
The $K_4^\perp$ term contains eight Bose operators, which are decoupled into
four according to
\ba
\lefteqn{
- \frac{K_{4,N}^\perp}{2N^3} \left[ (\sigma^z)_\beta^\alpha b_\beta^\ast
    b_\alpha \right]^4
\to \frac{N K_{4,N}^\perp}{2} R_z^2 } \nonumber \\
& & \qquad{}-
  \frac{K_{4,N}^\perp}{N} R_z \left[ (\sigma^z)_\beta^\alpha b_\beta^\ast
    b_\alpha \right]^2 .
\ea
To make the transformation well defined, $R_z$ is to be
integrated from $-\infty$ to $+\infty$ if $K_4^\perp > 0$
but from $-i\infty$ to $+i\infty$ if $K_4^\perp < 0$.
The remaining four-boson term is grouped together with the $K_2$ term
and decoupled,
\ba
\lefteqn{
- \frac{K_{2,N}+2 K_{4,N}^\perp R_z}{2N} \left[
  (\sigma^z)_\beta^\alpha b_\beta^\ast b_\alpha \right]^2
  } \nonumber \\
& & \qquad\to
  \frac{N(K_{2,N}+2 K_{4,N}^\perp R_z)}{2} P_z^2
  \nonumber \\
& & \qquad\quad{}- (K_{2,N}+2K_{4,N}^\perp R_z) P_z
  (\sigma^z)_\beta^\alpha b_\beta^\ast
    b_\alpha .
\ea
Here, the direction of integration of $P_z$ must be chosen such that
$(K_{2,N}+2 K_{4,N}^\perp R_z) P_z^2$ is real and positive at the end points.
The decoupling of the $K_4^\|$ terms is done similarly, introducing
additional fields $R_{x,y}$ and $P_{x,y}$.

Up to this point no approximation, apart from
the continuum limit, has been made.
Now we consider the mean-field approximation for this system.
This is exact for $N\to\infty$ but serves as a lowest-order
approximation for $N=2$. The mean-field theory is defined by
all auxiliary fields being constant in time. Since the system is
ferromagnetic, we assume that they are
constant in space as well and write them as $\overline\lambda$,
$\overline{\bf Q}$, $\overline{\bf P}$, and $\overline{\bf R}$.
Then, the action is
bilinear in the fields $b_\alpha$. Written in terms of Fourier
transforms the mean-field partition function reads
\be
Z_0 = Z_c\,
  \int \!\! \prod_{\alpha,{\bf k},i\omega_n} \!\!
    d^2b_\alpha({\bf k},i\omega_n)
  \: \exp\!\bigg(- \int d^2k \sum_{i\omega_n} {\cal L}_0[b] \bigg)
\ee
with the $b_\alpha$-independent part
\ba
Z_c & = & \exp\!\bigg(\!\!
  - {\cal N}N\beta J \overline{\bf Q}\cdot\overline{\bf Q} a^2
  + {\cal N}a^2NS\beta\overline{\lambda}
  \nonumber \\
& & {}- {\cal N}N\frac{\beta \tilde K_2}{2}\,\overline{P}_z^2
  - {\cal N}N\frac{\beta \tilde K_4^\perp}{2}\,
    \overline{R}_z^2
  \nonumber \\
& & {}- {\cal N}N \beta \tilde K_4^\perp\,
    \overline{R}_z \overline{P}_z^2
  - {\cal N}N\frac{\beta \tilde K_4^\|}{2} \left[
   \overline{R}_x^2 + \overline{R}_y^2 \right]
  \nonumber \\
& & {}- {\cal N}N \beta \tilde K_4^\|
  \times \left[
   \overline{R}_x \overline{P}_x^2 + \overline{R}_y \overline{P}_y^2
   \right] \bigg) ,
\ea
where ${\cal N}$ is the number of sites. The Lagrangian is
\ba
\lefteqn{
{\cal L}_0[b] = \beta a^2 \sum_\alpha
  \Big(\!-i\hbar\omega_n+J S k^2a^2-\frac{\beta J}{S}\,
    \overline{\bf Q}\cdot\overline{\bf Q}a^2
  } \nonumber \\
& & \quad\:{}+ a^2\overline{\lambda} \Big) \,
  b_\alpha^\ast({\bf k},i\omega_n) b_\alpha({\bf k},i\omega_n)
  \nonumber \\
& & \quad{}- \beta a^2 \sum_{\alpha\beta}
  \bigg( \frac{{\bf B}}{2}\cdot\text{\boldmath$\sigma$}_\beta^\alpha
    + \tilde K_2\, \overline{P}_z (\sigma^z)_\beta^\alpha
  \nonumber \\
& & \quad\:{}+ 2\,\tilde K_4^\perp\,
  \overline{R}_z \overline{P}_z (\sigma^z)_\beta^\alpha
  \nonumber \\
& & \quad\:{}+ 2\,\tilde K_4^\|
    \left[\overline{R}_x \overline{P}_x
    (\sigma^x)_\beta^\alpha
    + \overline{R}_y \overline{P}_y (\sigma^y)_\beta^\alpha \right]
  \! \bigg)\, b_\beta^\ast b_\alpha .
\ea
After integrating out the $b_\alpha$ fields and performing the sum over
field indices $\alpha$, we obtain, up to an irrelevant constant factor,
\be
Z_0 = Z_c\, \exp\!\bigg( -\frac{N}{2}\,\frac{{\cal N}a^2}{4\pi^2} \int d^2k
  \sum_{i\omega_n} \ln\det
  \text{\sf M}({\bf k},i\omega_n) \bigg)
\label{2.Z02}
\ee
with the $2\times2$ matrix
\ba
{\sf M} & = &
  (-i\beta\hbar\omega_n + \beta JSk^2a^2+\overline{\Lambda})
  \text{\bf 1}
  - \frac{\beta\bf B}{2} \cdot\text{\boldmath$\sigma$}
  \nonumber \\
& & {}- \beta \tilde K_2\, \overline{P}_z \sigma^z
  - 2\,\beta \tilde K_4^\perp\,
    \overline{R}_z \overline{P}_z \sigma^z
  \nonumber \\
& & {}- 2\,\beta \tilde K_4^\|
    (\overline{R}_x \overline{P}_x \sigma^x
    + \overline{R}_y \overline{P}_y \sigma^y) .
\ea
Note that $Z_0$ only depends on $\overline{\lambda}$ and
$\overline{\bf Q}$ through
\be
\overline{\Lambda} \equiv a^2\beta\overline{\lambda}
  - \frac{\beta J}{S}\,\overline{\bf Q}\cdot\overline{\bf Q} a^2 ,
\ee
just as in the isotropic case.\cite{TGHS}
Evaluation of the frequency sum and momentum integral in Eq.~(\ref{2.Z02})
yields the free energy per boson
\ba
\beta F_0 & \equiv & -\frac1{{\cal N}N}\,\ln Z_0
  \nonumber \\
& = & - S\overline{\Lambda}
    + \frac{\beta \tilde K_2}{2}\, \overline{P}_z^2
    + \frac{\beta \tilde K_4^\perp}{2}\,
   \overline{R}_z^2
  \nonumber \\
& & \quad{}+ \beta \tilde K_4^\perp\,
      \overline{R}_z \overline{P}_z^2
    + \frac{\beta \tilde K_4^\|}{2}
      \left( \overline{R}_x^2 + \overline{R}_y^2 \right)
  \nonumber \\
& & \quad{}
    + \beta \tilde K_4^\|
      \left( \overline{R}_x \overline{P}_x^2
      + \overline{R}_y \overline{P}_y^2 \right)
  \nonumber \\
& & \quad{}+ \frac{1}{8\pi\beta JS}\, \Big[
      \text{diln} \left(1-e^{-\overline{\Lambda}+\beta B'/2}\right)
  \nonumber \\
& & \qquad{}+ \text{diln} \left(1-e^{-\overline{\Lambda}-\beta B'/2}\right)
  \Big]
\ea
with $B'=|{\bf B}'|$ and
\ba
B'_x & = & B_x + 4\, \tilde K_4^\|\,
  \overline{R}_x \overline{P}_x , \\
B'_y & = & B_y + 4\, \tilde K_4^\|\,
  \overline{R}_y \overline{P}_y , \\
B'_z & = & B_z + 2\, \tilde K_2\, \overline{P}_z
    + 4\, \tilde K_4^\perp\, \overline{R}_z \overline{P}_z .
\ea
The dilogarithm function is defined by the series representation
\be
\text{diln}\,x \equiv -\sum_{n=1}^\infty \frac{(1-x)^n}{n^2} .
\ee
We are mainly interested in the magnetization
\ba
\lefteqn{
{\bf M}_0 \equiv -2\,\frac{d F_0}{d{\bf B}}
  = -2\,\frac{\partial F_0}{\partial{\bf B}}
  = -\frac1{8\pi\beta JS}\,\frac{{\bf B}'}{B'}
  } \nonumber \\
& & \quad\times
  \left[\ln\left(1-e^{-\overline{\Lambda}+\beta B'/2}\right)
      - \ln\left(1-e^{-\overline{\Lambda}-\beta B'/2}\right) \right] .
\ea
We have used that the derivatives of $F_0$ with respect to
$\overline{\Lambda}$ and $\overline{\bf P}$ vanish at the
saddle point. A factor of $2$ appears, since in the physical system
two bosons make up one spin.
For the isotropic case ${\bf B}'$ equals ${\bf B}$ and we recover the
results of Refs.~\onlinecite{RS} and \onlinecite{TGHS}.

Finally, we have to determine the auxiliary fields by finding the lowest
stable free-energy saddle point. We set the derivatives
of $\ln Z_0$ with respect to $\overline{\Lambda}$, $\overline{\bf P}$,
and $\overline{\bf R}$ to zero and solve the resulting equations
simultaneously. From $\partial\ln Z_0/\partial\overline{\Lambda}=0$
we obtain the saddle-point equation
\ba
S & = & -\frac1{8\pi\beta JS}
  \Big[\ln\left(1-e^{-\overline{\Lambda}+\beta B'/2}\right)
  \nonumber \\
& & {}+ \ln\left(1-e^{-\overline{\Lambda}-\beta B'/2}\right) \Big] ,
\ea
which can be analytically solved for $\Lambda$.\cite{TGHS}
By taking the derivatives with respect to the other auxiliary fields one
finds that for all saddle-point solutions
\be
\overline{R}_i=\overline{P}_i^2, \quad i=x,y,z .
\label{a.RP1}
\ee
This equation either follows directly from the saddle-point equations
or, for special values of $\overline{R}_i$, the free energy is found
to be independent of the corresponding $\overline{P}_i$ so that we can
choose $\overline{P}_i$ to satisfy Eq.~(\ref{a.RP1}).
We end up with a free energy which depends on the four
auxiliary parameters $\overline{\Lambda}$ and $\overline{\bf P}$,
\ba
\beta F_0
  & \!=\! & - S\overline{\Lambda}
    + \frac{\beta \tilde K_2}{2}\, \overline{P}_z^2
  \nonumber \\
& & {}+ \frac{3\beta \tilde K_4^\perp}{2}\, \overline{P}_z^4
    + \frac{3\beta \tilde K_4^\|}{2}\,
      \left( \overline{P}_x^4 + \overline{P}_y^4 \right)
  \nonumber \\
& & {}+ \frac{1}{8\pi\beta JS}\, \Big[
      \text{diln} \left(1-e^{-\overline{\Lambda}+\beta B'/2}\right)
  \nonumber \\
& & \quad{}+ \text{diln} \left(1-e^{-\overline{\Lambda}-\beta B'/2}\right)
  \Big]
\ea
with
\ba
B'_x & = & B_x + 4\, \tilde K_4^\|\, \overline{P}_x^3 , \\
B'_y & = & B_y + 4\, \tilde K_4^\|\, \overline{P}_y^3 , \\
B'_z & = & B_z + 2\, \tilde K_2\, \overline{P}_z
     + 4\, \tilde K_4^\perp\, \overline{P}_z^3 .
\ea
In the uniaxial case $K_4^\|=0$ there are only two relevant parameters,
$\overline{\Lambda}$ and $\overline{P}_z$.

By solving the saddle-point equations
$\partial\ln Z_0/\partial \overline{P}_i=0$ we find that there are
typically two solutions for each component $\overline{P}_i$.
One is $\overline{P}_i=M_{0,i}$, {\it i.e.},
$\overline{P}_i$ is the corresponding magnetization
component, and the other is given by $B'_i=B_i$. For example,
in the uniaxial case, $K_4^\|=0$ ($\tilde K_4^\|=0$)
the free energy does not depend on $\overline{P}_x$ and $\overline{P}_y$
whereas $M_{0,x}$ and $M_{0,y}$ have well-defined values,
which need not be zero in the presence of an external field.
Thus $\overline{P}_{x,y}$ cannot equal $M_{0,x,y}$. The
saddle-point equations are solved by the second type of solution,
$B'_{x,y}=B_{x,y}$, since $K_4^\perp=0$. On the other hand,
the $z$ component is $\overline{P}_z = M_{0,z}$.

It turns out that the relevant saddle point of the free energy $F_0$
is not a minimum with respect to all four parameters. For example, it
is always a maximum with respect to $\overline{\Lambda}$ (or
$\overline{\lambda}$). Intuitively, one would expect $\overline{\Lambda}$
to run off to infinity in this case. However, since $\lambda$ in
Eq.~(\ref{2.Zf1}) has to be integrated from $-i\infty$ to $+i\infty$,
fluctuations in $\lambda$ are along the imaginary direction,
$\lambda({\bf r},\tau) = \overline{\lambda} + i\Delta\lambda({\bf r},\tau)$,
where $\Delta\lambda$ is real. In particular, this holds for global
changes of $\overline{\lambda}$. With respect to $\Delta\lambda$, the free
energy indeed has a minimum at $\Delta\lambda=0$. For the
Hubbard-Stratonovich fields $\overline{\bf P}$ and $\overline{\bf R}$
the direction of fluctuations depends on the values of the anisotropy
constants. The stability of each saddle point against global changes
of the auxiliary parameters has to be checked in order
to find the lowest stable one. Note that for positive anisotropy constants
the fluctuations of $\overline{\bf P}$ and $\overline{\bf Q}$
are purely real and the stability analysis becomes
straightforward.

%%% FIGURES

\newpage

\newpage

\begin{thebibliography}{99}

\bibitem[*]{email}Electronic address: {\tt timm@physik.fu-berlin.de}.

\bibitem{Schwinger}J. Schwinger, {\it Quantum Theory of Angular
Momentum}, edited by L. Biedenharn and H. Van Dam (Academic, New York,
1965).

\bibitem{RN}N. Read and D.M. Newns, J. Phys.\ C {\bf 16}, 3273
(1983).

\bibitem{AA}D.P. Arovas and A. Auerbach, Phys.\ Rev.\ B {\bf 38},
316 (1988); A. Auerbach and D.P. Arovas, Phys.\ Rev.\ Lett.\ {\bf 61},
617 (1988).

\bibitem{Auerbach}A. Auerbach, {\it Interacting Electrons and
Quantum Magnetism\/} (Springer, New York, 1994).

\bibitem{Starykh}O.A. Starykh, Phys.\ Rev.\ B {\bf 50}, 16428 (1994);
A.V. Chubukov and O.A. Starykh, {\it ibid}.\ {\bf 52}, 440 (1995).

\bibitem{RS}N. Read and S. Sachdev, Phys.\ Rev.\ Lett. {\bf 75},
3509 (1995).

\bibitem{TGHS}C. Timm, S.M. Girvin, P. Henelius, and A.W. Sandvik,
Phys.\ Rev.\ B {\bf 58}, 1464 (1998).

\bibitem{MW}N.D. Mermin and H. Wagner, Phys.\ Rev.\ Lett.\ {\bf 17},
1133 (1966).

%%\bibitem{RStopo}N. Read and S. Sachdev, Nucl.\ Phys.\ B {\bf 316}, 609
%%(1989); Phys.\ Rev.\ Lett.\ {\bf 62}, 1694 (1989).

\bibitem{HP}T. Holstein and H. Primakoff, Phys.\ Rev.\ {\bf 58},
1908 (1940).

\bibitem{ex2a}D.P. Pappas, K.P. K\"amper, and H. Hopster, Phys.\ Rev.\ Lett.\
{\bf 64}, 3179 (1990).

\bibitem{ex2b}J. Thomassen, F. May, B. Feldmann, M. Wuttig, and
H. Ibach, Phys.\ Rev.\ Lett.\ {\bf 69}, 3831 (1992).

\bibitem{ex2c}U. Gradmann, {\it Handbook of Magnetic Materials\/}
(Elsevier, Amsterdam, 1993), Vol. 7; J.A.C. Bland and B. Heinrich,
{\it Ultrathin Magnetic Structures\/} (Springer Verlag, Berlin, 1994).

\bibitem{ex2d}Z.Q. Qiu, J. Pearson, and S.D. Bader, Phys.\ Rev.\ Lett.\
{\bf 70}, 1006 (1993);
D. Li, M. Freitag, J. Pearson, Z.Q. Qiu, and S.D. Bader,
{\it ibid}.\ {\bf 72}, 3112 (1994).

\bibitem{ex2e}G. Lugert, W. Robl, L. Pfau, M. Brockmann, and G. Bayreuther,
J. Magn.\ Magn.\ Mater.\ {\bf 121}, 498 (1993).

\bibitem{FMR}B. Schulz and K. Baberschke, Phys.\ Rev.\ B {\bf 50},
13467 (1994); M. Farle, W. Platow, A.N. Anisimov, B. Schulz,
and K. Baberschke, J. Magn.\ Magn.\ Mater.\ {\bf 165}, 74 (1997).

\bibitem{Bab97}K. Baberschke, Appl.\ Phys.\ A {\bf 62}, 417 (1996);
M. Farle, B. Mirwald-Schulz, A.N. Anisimov, W. Platow,
and K. Baberschke, Phys.\ Rev.\ B {\bf 55}, 3708 (1997).

\bibitem{ex2f}R. Allenspach, J. Magn.\ Magn.\ Mater.\ {\bf 129}, 160 (1994).

\bibitem{ex2g}O. Schulte, F. Klose, and W. Felsch, Phys.\ Rev.\ B {\bf 52},
6480 (1995).

\bibitem{ex2h}A. Berger and H. Hopster, Phys.\ Rev.\ Lett.\ {\bf 76},
519 (1996).

\bibitem{ex2i}G. Garreau, E. Beaurepaire, K. Ounadjela, and M. Farle,
Phys.\ Rev.\ B {\bf 53}, 1083 (1996).


\bibitem{Kit51}C. Herring and C. Kittel, Phys.\ Rev.\ {\bf 81}, 869
(1951); S.V. Maleev, Sov.\ Phys.\ JETP {\bf 43}, 1240 (1976);
V.L. Pokrovsky and M.V. Feigel'man, Sov.\ Phys.\ JETP {\bf 45}, 291 (1977).

\bibitem{Mil91}R.P.Erickson and D.L. Mills, Phys.\ Rev.\ B {\bf 43}, 10715
(1991); {\bf 44}, 11825 (1991).

\bibitem{Tya59}S.V. Tyablikov, Ukr.\ Mat.\ Zh.\ {\bf 11} 287 (1959).

\bibitem{Tha62}R. Tahir-Kheli and D. ter Haar, Phys.\ Rev.\ {\bf 127},
88 (1962).

\bibitem{Cal63}H.B. Callen, Phys.\ Rev.\ {\bf 130}, 890 (1963).

\bibitem{Tya67}S.V.Tyablikov, {\it Methods in the quantum theory of
magnetism\/} (Plenum Press, New York, 1967).

\bibitem{FJK99}P. Fr\"obrich, P.J. Jensen, and P.J. Kuntz,
Europ.\ Phys.\ J. B {\bf 13}, 477 (1999).

\bibitem{Doe64}W. D\"oring, Z. Naturforsch.\ {\bf A16}, 1008 (1961).

\bibitem{Hau72}W. Haubenreisser, W. Brodkorb, A. Corciovei, and
G. Costache, Phys.\ Stat.\ Solidi B {\bf 53}, 9 (1972).

\bibitem{Die79}Diep-the-Hung, J.C.S. Levy, and O. Nagai,
Phys.\ Stat.\ Solidi B {\bf 93}, 351 (1979).

\bibitem{Chi90}L.P. Shi and W.G. Yang, J. Phys.\ Condens.\ Mater.\
{\bf 4}, 7997 (1992).

\bibitem{Bru91}P. Bruno, Phys.\ Rev.\ B {\bf 43}, 6015 (1991).

\bibitem{Mor94}D.K. Morr, P.J. Jensen, and K.H. Bennemann,
Surf.\ Sci.\ {\bf 307--309}, 1109 (1994).

\bibitem{HSTG}P. Henelius, A.W. Sandvik, C. Timm, and S.M. Girvin,
Phys.\ Rev.\ B {\bf 61}, 364 (2000).

\bibitem{EFJK}A. Ecker, P. Fr\"obrich, P.J. Jensen, and P.J. Kuntz,
J. Phys.\ Cond.\ Mat.\ {\bf 11}, 1557 (1999).

\bibitem{Erd75}S.B. Haley, Phys.\ Rev.\ B {\bf 17}, 337 (1978).

\bibitem{And64}F.B. Anderson and H.B. Callen, Phys.\ Rev.\ {\bf 136},
A1068 (1964).

\bibitem{Lin67}M.E. Lines, Phys.\ Rev.\ {\bf 156}, 534 (1967).

\bibitem{Jen96}P.J. Jensen and K.H. Bennemann, Solid State
Commun.\ {\bf 100}, 585 (1996).

\bibitem{Mil95}Y. Millev and M. F\"ahnle, Phys.\ Rev.\ B {\bf 51},
2937 (1995).

\bibitem{Jen98}P.J. Jensen and K.H. Bennemann, Solid State
Commun.\ {\bf 105}, 577 (1998), and references therein.

\bibitem{JeB98}P.J. Jensen and K.H. Bennemann, {\it Magnetism and
Electronic Correlations in Local-Moment Systems: Rare-Earth Elements
and Compounds}, edited by M. Donath, P.A. Dowben, and
W. Nolting (World Scientific, Singapore, 1998), p.\ 113.

\bibitem{Mos94}A. Moschel and K.D. Usadel, Phys.\ Rev.\ B {\bf 49},
12868 (1994).

\bibitem{Huc97}A. Hucht and K.D. Usadel, Phys.\ Rev.\ B {\bf 55}, 12309
(1997); Phil.\ Mag.\ B {\bf 80}, 275 (2000).

\bibitem{Jen93}P.J. Jensen and K.H. Bennemann, Ann.\ Physik (Leipzig)
{\bf 2}, 475 (1993).

\bibitem{cont}X. Hu and Y. Kawazoe, Phys.\ Rev.\ B {\bf 54}, 65 (1996),
and references therein.

\bibitem{renorm}P. Politi, A. Rettori, M.G. Pini, and D. Pescia,
J. Magn.\ Magn.\ Mater.\ {\bf 140-144}, 647 (1995);
A. Abanov, V. Kalatsky, V.L. Pokrovsky, and W.M. Saslow,
Phys.\ Rev.\ B {\bf 51}, 1023 (1995), and references therein.

\bibitem{Cas59}H.B.G. Casimir, J. Smit, U. Enz, J.F. Fast, H.P.J. Wijn,
E.W. Gorter, A.J.W. Duyvesteyn, J.D. Fast, and J.J. de Jong,
J. Phys. Radium {\bf 20}, 360 (1959).

\bibitem{MOK}Y.T. Millev, H.P. Oepen, and J. Kirschner,
Phys.\ Rev.\ B {\bf 57}, 5837 (1998); {\bf 57}, 5848 (1998).

\bibitem{MC}S.T. Chui, Phys.\ Rev.\ B {\bf 50}, 12559 (1994);
A. Hucht, A. Moschel, and K.D. Usadel, J. Magn.\ Magn.\ Mater.\
{\bf 148}, 32 (1995); 
O. Iglesias, A. Valencia, and A. Labarta, {\it ibid}.\
{\bf 196-197}, 819 (1999).

\bibitem{QHferro1}S.L. Sondhi, A. Karlhede, S.A. Kivelson, and
E.H. Rezayi, Phys. Rev. B {\bf 47}, 16419 (1993).

\bibitem{QHferro2}R. C\^ot\'e, A.H. MacDonald, L. Brey, H.A. Fertig,
S.M. Girvin, and H.T.C. Stoof, Phys.\ Rev.\ Lett.\ {\bf 78}, 4825
(1997); S.M. Girvin and A.H. MacDonald, in {\it Perspectives
in Quantum Hall Effects}, edited by S. Das Sarma and A. Pinczuk (Wiley,
New York, 1997).

\bibitem{Manfra}M.J. Manfra, E.H. Aifer, B.B. Goldberg, D.A. Broido,
L. Pfeiffer, and K. West, Phys.\ Rev.\ B {\bf 54}, R17327 (1996).

\bibitem{rem:QHdip}The dipolar interaction may be important, though.
Because of the large external magnetic fields present in quantum
Hall experiments, we do not expect the dipolar interaction to lead to the
formation of domains. However, it does favour a spin orientation
parallel to the 2D electron gas. This effect may be
observable in experiments with magnetic fields tilted away from the
normal axis.

\bibitem{newMillev}Y. Millev and M. F\"ahnle, Phys.\ Rev.\ B {\bf 52},
4336 (1995).

\bibitem{Jenxx}P.J. Jensen (unpublished).

\bibitem{rem:Qspin}At first glance it may seem surprising that a factor
of $S^2$ instead of the quantum-mechanical
$S(S+1)$ appears in the expressions. However, the
factor comes from the exchange interaction, as indicated by the factor
$J$. The exchange term contains two spin operators at different sites,
which commute. Thus ${\bf S}_i\cdot{\bf S}_j = S^2$ if
${\bf S}_i\|{\bf S}_j$.

\bibitem{Jen90}P.J. Jensen and K.H. Bennemann,
Phys.\ Rev.\ B {\bf 42}, 849 (1990).

\bibitem{rem:domains}The true ground state of a perpendicularly
magnetized thin film is not a homogeneous one but, due to the long-range
dipole coupling, a stripe domain
phase with a vanishing net magnetization. By considering this multidomain
state the phase diagram of Fig.~\ref{fig3} is altered somewhat. For a
thorough discussion of these phases see Ref.~\onlinecite{JeB95}. In the
present work we assume homogeneous phases.

\bibitem{rem:spin}The spin quantum number does not affect the
qualitative form of the phase diagram, as long as $S\ge 2$ so that
fourth-order anisotropies are relevant. We here study $S=2$.

\bibitem{cata}P.T. Saunders, {\it An introduction to catastrophy
theory\/} (Cam\-bridge University Press, Cambridge, 1980).

\bibitem{Cal66}H.B. Callen and E.R. Callen, J. Phys.\ Chem.\ Solids
{\bf 27}, 1271 (1966).

\bibitem{MHB96}T.H. Moos, W. H\"ubner, and K.H. Bennemann, Solid
State Commun.\ {\bf 98}, 639 (1996).

\bibitem{JeB95}P.J. Jensen and K.H. Bennemann,
Phys.\ Rev.\ B {\bf 52}, 16012 (1995), and references therein.


\end{thebibliography}
\end{document}